\documentclass[a4paper]{article}

\usepackage[latin1]{inputenc}
\usepackage{amsfonts}
\usepackage{amsmath}
\usepackage{amssymb}
\usepackage[UKenglish]{babel}
\usepackage{fontenc}
\usepackage{graphicx}
\usepackage{amscd}
\usepackage{psfrag}
\usepackage{amscd}
\usepackage{amsthm}

\newcommand{\CC}{\ensuremath{\mathbb{C}}}
\newcommand{\RR}{\ensuremath{\mathbb{R}}}
\newcommand{\NN}{\ensuremath{\mathbb{N}}}
\newcommand{\ZZ}{\ensuremath{\mathbb{Z}}}

\newcommand{\CP}[1]{\ensuremath{\mathbb{CP}^{#1}}}

\renewcommand{\d}{\ensuremath{\mathrm{d}}}

\newcommand{\modsq}[1]{\ensuremath{|#1|^2}}

\newcommand{\mmodsq}[1]{\ensuremath{\|#1\|^2}}
\newcommand{\Mmodsq}[1]{\ensuremath{\left\|#1\right\|^2}}

\newcommand{\SU}[1]{\ensuremath{\mathrm{SU}(#1)}}

\newcommand{\U}[1]{\ensuremath{\mathrm{U}(#1)}}
\renewcommand{\u}[1]{\ensuremath{\mathfrak{u}(#1)}}

\newcommand{\rk}{\ensuremath{\mathrm{rank}}}

\newcommand{\im}{\ensuremath{\bold{i}}}
\newcommand{\half}{\ensuremath{\frac{1}{2}}}
\newcommand{\quart}{\ensuremath{\frac{1}{4}}}
\newcommand{\ihalf}{\ensuremath{\frac{\im}{2}}}

\newcommand{\w}{\wedge}

\newcommand{\tr}[1]{\ensuremath{\mathrm{tr} \! \left( #1 \right)} }

\newcommand{\dvol}{\ensuremath{\mathrm{dvol}}}

\newcommand{\al}{\alpha}
\newcommand{\be}{\beta}

\newcommand{\vol}[1]{\ensuremath{\mathrm{Vol}(#1)}}

\newcommand{\curlF}{\ensuremath{\mathcal{F}}}

\newcommand{\curlE}{\ensuremath{\mathcal{E}}}

\newcommand{\Hc}{\ensuremath{\mathrm{H}}}

\newcommand{\pd}{\ensuremath{\partial}}

\newcommand{\dgr}{\dagger}

\newcommand{\ph}{\phantom}

\newcommand{\lpp}{\ensuremath{(\!(}}
\newcommand{\rpp}{\ensuremath{)\!)}}

\newcommand{\C}{\ensuremath{\textit{C}}}
\newcommand{\ch}{\ensuremath{\textit{ch}}}
\newcommand{\Ch}{\ensuremath{\textit{Ch}}}

\begin{document}

\pagestyle{plain}

\title{
\vskip -70pt
\begin{flushright}
{\normalsize DAMTP-2014-18} \\
\end{flushright}
\vskip 50pt
{\bf \Large A ladder of topologically non-trivial non-BPS states}
\vskip 30pt
}

\author{
Daniele Dorigoni\footnote{dd365@damtp.cam.ac.uk}\,\, and Norman A. Rink\footnote{nar43@cantab.net} \\ \\
{\sl Department of Applied Mathematics and Theoretical Physics,}\\
{\sl University of Cambridge,}\\
{\sl Wilberforce Road, Cambridge CB3 0WA, England.}\\
}

\vskip 20pt
\date{30 March 2014}
\vskip 20pt

\maketitle

\begin{abstract}

We consider a simple quiver gauge theory with gauge group $\U{r_1}\times \U{r_2}$ and a Higgs field in the bi-fundamental representation. The background for this theory is  a compact K\"ahler manifold $M$. 
For a careful but natural choice of Higgs field potential the second order field equations can be replaced with a set of first order BPS equations. 
We show that the theory admits two energy gaps: The vacuum is topologically trivial but has finite, non-zero energy and is not a BPS state. The second gap lies between the vacuum and the first BPS state. In this gap we find a ladder of states with non-trivial topology, at equidistant energy levels.
We give a semi-explicit construction for such topologically non-trivial non-BPS states.

\end{abstract}

\newpage

\newpage

\section{Introduction}

Many interesting field theories admit a special class of solutions, so-called BPS states. BPS states are special in that they satisfy a set of first order field equations, which imply the usual second order equations of the field theory. Solutions of the first oder equations minimize the static energy functional in a fixed topological sector. In theories with BPS states the following statements usually hold: 
\begin{enumerate}
	\item The vacuum of the theory is a BPS state in the topologically trivial sector.
	\item The vacuum has zero energy.
	\item The first topologically non-trivial solution is a BPS state.
	\item In each topological sector the energy is minimized by a BPS state.
\end{enumerate}
The subject of this report is a simple quiver gauge theory which admits BPS states but for which none of the above statements hold.

More specifically, we study a theory with a single Higgs field $\phi$ that is charged under two gauge groups, $\U{r_1}$ and $\U{r_2}$. The background for this theory is a compact K\"ahler manifold $M$, with fixed area $\vol{M}$. We will see that this theory exhibits two energy gaps: First, the vacuum has non-zero energy, proportional to $\vol{M}$. While the vacuum has trivial topology, it is, however, not a BPS state. Second, between the vacuum and the lowest energy BPS state there is room for a \textit{ladder} of topologically non-trivial solutions at equidistant energy levels.
  
Our theory is not new: It is a standard example of a field theory that accommodates non-abelian vortices \cite{Bradlow:GP:stable, Popov:quiver, Popov:nonab, Manton:Rink:nonab}. In fact, a special case of our theory was studied in \cite{Manton:Rink:nonab} in the case where $M$ has complex dimension one. It was already observed in \cite{Manton:Rink:nonab} that this theory has two energy gaps, but the question whether there are solutions between those two gaps was left unanswered. Here we finally give this answer, and we put our analysis on a more general footing, relying on classical methods from complex geometry.

Although this report focuses on one specific theory, it is to be expected that similar results hold for a whole class of quiver gauge theories. This is because quiver gauge theories on $M$, with various numbers of Higgs fields, can be derived by dimensional reduction from pure Yang--Mills theory on $M\times\CP{1}$ \cite{Popov:quiver}. Properties of a special quiver gauge theory, like the one that is the subject of this report, must have their roots in the higher-dimensional Yang--Mills theory. Therefore other quiver guage theories, obtained by different ways of reducing dimensionally, must also reflect those properties.

Crucial to our observations is the following quartic potential for the Higgs field $\phi$,
\begin{equation}
V(\phi) = \quart \int_M \left( \modsq{\phi\phi^\dgr - \tau_1 I_{r_1}} + \modsq{\phi^\dgr\phi - \tau_2 I_{r_2}} \right)\dvol_M,
\end{equation}
where $I_{r_1}, I_{r_2}$ are the identity matrices of ranks $r_1, r_2$ respectively, and $\tau_1, \tau_2\in\RR$ are parameters that determine the self-coupling of the Higgs field. Quartic potentials as the one above are typical of theories that support vortices \cite{Bogomolny, Nielsen:Olesen, Manton:Sutcliffe}. If $V(\phi)$ is obtained by dimensionally reducing Yang--Mills theory on $M\times\CP{1}$, then $\tau_1$ and $\tau_2$ are constrained \cite{Popov:quiver, Manton:Rink:nonab}. Another way to derive the above expression for $V(\phi)$ is by integrating the so-called $D$-term equation in a supersymmetric version of our model. Corresponding to $\U{r_1}$ and $\U{r_2}$ there are two gauge potentials $A_1$ and $A_2$, whose supermultiplets contain the Lie algebra-valued scalar fields $D_1,D_2$ respectively. In terms of those scalar fields the supersymmetric potential $V(\phi)$ reads
\begin{equation}
V(\phi)= \int_M \left(  (D_1, - D_1 + \phi\phi^\dgr - \tau_1 I_{r_1}) + (D_2, -D_2 + \phi^\dgr \phi - \tau_2 I_{r_2}) \right) \dvol_M. \label{eq:D:term}
\end{equation} 
Note that since the gauge group contains two $\U{1}$ factors, the Fayet-Illiopulos terms $\tau_1 D_1$ and $\tau_2 D_2$ do not spoil supersymmetry. The fields $D_1,D_2$ are non-dynamical, and hence can be integrated out, leading to our original version of $V(\phi)$.

This report is organised as follows: in Section \ref{sec:notation} we review concepts and results from complex geometry, which will enable our analyses in subsequent sections. We formally introduce our quiver gauge theory by giving its static energy functional in Section \ref{sec:theory}, where we also derive the BPS equations and lower bounds for the energy functional. We introduce special equations for non-BPS solutions and study their properties in Section \ref{sec:non-BPS}.

\section{Complex and K\"ahler geometry} \label{sec:notation}

The quiver gauge theory we study in this report is set on a fixed K\"ahler background $M$. In this preliminary Section we review basic properties of K\"ahler manifolds and operators on them. This mainly serves to set up notation. We also introduce notation regarding vector bundles in the context of gauge theories. Useful references for this Section are standard textbooks on complex and K\"ahler geometry, such as \cite{Griffiths:Harris, Kobayashi:vector:bundles, Huybrechts}.

\subsection{K\"ahler manifolds and forms}

Let $M$ denote a compact K\"ahler manifold, with K\"ahler form $\omega$, and let $d$ be the complex dimension of $M$. We denote the space of (complex-valued) $k$-forms on $M$ as $\Omega^k(M)$. Since $M$ is in particular a complex manifold, forms decompose into their holomorphic and anti-holomorphic parts, i.e.
\begin{align}
	\Omega^k(M) = \bigoplus_{p+q=k} \Omega^{p,q}(M). \label{eq:form:decomp}
\end{align}
By definition of the K\"ahler form, $\omega\in\Omega^{1,1}(M)$. The volume form of $M$ is given by
\begin{align}
	\dvol_M = \omega^{[d]} \equiv \frac{\omega^d}{d!},
\end{align}
where the right identity should be regarded as a definition of the superscript $[\:\:]$.

Let $(\,,\,)$ be the standard scalar product between $(p,q)$-forms, i.e.
\begin{align}
	(\,,\,) \colon \Omega^{p,q}(M)\times \Omega^{p,q}(M) \to \Omega^{0}(M),
\end{align}
and $(\,,\,)$ is $\CC$-linear in its first argument and $\CC$-anti-linear in its second. We use this scalar product to define the Hodge $*$ operator,
\begin{align}
	&\nonumber* \colon \Omega^{p,q}(M) \to \Omega^{d-q, d-p}(M), \\
	&(\al, \be)\, \dvol_M = \al \w * \bar\be, \quad \al, \be\in\Omega^{p,q}(M).
\end{align}
We also introduce the scalar product $\lpp\, , \,\rpp$ on $(p,q)$-forms,
\begin{align}
	&\nonumber\lpp\,,\,\rpp \colon \Omega^{p,q}(M)\times \Omega^{p,q}(M) \to \CC, \\
	&\lpp \al, \be \rpp = \int_M (\al, \be)\, \dvol_M, \quad \al, \be\in\Omega^{p,q}(M),
\end{align}
and we use the short-hand notation
\begin{align}
	\modsq{\al} = (\al, \al), \quad \mmodsq{\al} = \lpp \al, \al\rpp, \quad \al \in\Omega^{p,q}(M). \label{eq:short:hand}
\end{align}

A natural operator on the K\"ahler manifold $M$ is the Lefschetz operator $L$, given by
\begin{align}
	&\nonumber L \colon \Omega^{p,q}(M) \to \Omega^{p+1,q+1}(M)\,, \\
	&L\al = \omega\w\al, \quad \al\in\Omega^{p,q}(M).
\end{align}
The adjoint of $L$ with respect to the scalar product $\lpp\, , \,\rpp$ is denoted as $\Lambda$,
\begin{align}
	&\nonumber \Lambda \colon \Omega^{p,q}(M) \to \Omega^{p-1,q-1}(M) \\
	&(\Lambda\al, \be) = (\al, L\be), \quad \al\in\Omega^{p,q}(M), \be\in\Omega^{p-1,q-1}(M),
\end{align}
and from the previous line one can conclude that $\Lambda = *^{-1}\circ L\circ *$. Generally, for any operator on $ \Omega^{p,q}(M)$, we use the superscript $^*$ to denote its adjoint with respect to $\lpp\, , \,\rpp$.

Just as any $1$-form on $M$, the exterior derivative,
\begin{align}
	\d \colon \Omega^{k}(M) \to \Omega^{k+1}(M),
\end{align}
also decomposes into its $(1,0)$ and $(0,1)$ parts. That is, $\d = \pd + \bar\pd$, where
\begin{align}
	&\pd \colon \Omega^{p,q}(M) \to \Omega^{p+1,q}(M), \\
	&\bar\pd \colon \Omega^{p,q}(M) \to \Omega^{p,q+1}(M).
\end{align}

\subsection{Vector bundles and gauge theory} \label{sec:bundles}

For the purpose of introducing notation and reviewing the geometry of vector bundles, we let $E$ denote a complex vector bundle over $M$. Let $r$ be the rank of $E$. In the context of gauge theory, vector bundles are equipped with a hermitian structure. This allows us to choose unitary frames that locally span the fibre of $E$. Hence the structure group $G$ of $E$ can be reduced to a unitary group, i.e.~$G\subset\U{r}$.

In this subsection we denote as $D$ a covariant derivate on $E$ which is compatible with the hermitian structure. If $\phi$ is a section of $E$, then, on a local neighbourhood $U\subset M$,
\begin{align}
	D \phi = \d \phi + A\phi,
\end{align}
where $A$ is a Lie algebra valued 1-form on $U$, in symbols,
\begin{align}
	A \in \Omega^1(U, \u{r}). \label{eq:A:unitary}
\end{align}
In the usual terminology of gauge theory, $A$ is of course referred to as the local gauge potential. The corresponding field strength $F$ is 
\begin{align}
	F = D \circ D = \d A + A\w A \in \Omega^2(U, \u{r}),
\end{align}
and in geometric terms this is the curvature of $E$.

The curvature $F$ can be used to calculate topological invariants (see e.g.~\cite{Kobayashi:vector:bundles, Huybrechts, Nakahara}). In this report we shall have use for the first Chern class,
\begin{align}
	&\textit{c}_1(E) =  \frac{\im}{2\pi} [\tr{F}] \in \Hc^2(M,\RR),
\end{align}
and the second Chern character,
\begin{align}	
	&\ch_2(E) =  - \frac{1}{8\pi^2} [\tr{F\w F}] \in \Hc^4(M,\RR),
\end{align}
where the square brackets on the right-hand sides mean that the cohomology class of a closed form is taken. Cohomology classes like the above are generally referred to as characteristic classes of the bundle $E$. One obtains characteristic numbers by integrating over $M$,
\begin{align}
	&\C_1(E,\omega) = \int_M \textit{c}_1(E) \w \omega^{[d-1]}, \\
	&\Ch_2(E,\omega) = \int_M \ch_2(E)\w \omega^{[d-2]}, \label{eq:Chern:char}
\end{align}
cf.~\cite{Bradlow:line:bundles}. Since the K\"ahler form $\omega$ appears under the integrals, the numbers $\C_1(E,\omega)$, $\Ch_2(E,\omega)$ are not purely topological invariants but also depend on the geometry of $M$. However, since in this report we always assume $M$ and $\omega$ to be fixed, we adopt the lax terminology of referring to $\C_1(E,\omega)$, $\Ch_2(E,\omega)$ as topological terms.

We also use $(\,,\,)$ to denote the standard scalar product between Lie algebra valued $(p,q)$-forms. Since the Hodge $*$ operator extends to Lie algebra valued forms by acting trivially on the Lie algebra components, the following can be regarded as a definition of $(\,,\,)$:
\begin{align}
	&\nonumber (\,,\,) \colon \Omega^{p,q}(M, \u{r})\times \Omega^{p,q}(M, \u{r}) \to \Omega^{0}(M), \\
	&(\al, \be)\, \dvol_M = \tr{\al \w * \be^\dgr}.
\end{align}
As before we define $\lpp\, , \,\rpp$ by
\begin{align}
	&\nonumber \lpp\,,\,\rpp \colon \Omega^{p,q}(M, \u{r})\times \Omega^{p,q}(M, \u{r}) \to \CC, \\
	&\lpp \al, \be \rpp = \int_M (\al, \be)\, \dvol_M,
\end{align}
and we use the same short-hand notation as in \eqref{eq:short:hand}. 

Lastly we note that the covariant derivate $D$ also has a decomposition according to \eqref{eq:form:decomp},
\begin{align}
	D = D^{1,0} + D^{0,1},
\end{align}
where locally, i.e.~on a neighbourhood $U\subset M$,
\begin{align}
	&D^{1,0} = \pd        + A^{1,0}, \quad  A^{1,0} \in \Omega^{1,0}(U, \u{r}), \\
	&D^{0,1} = \bar\pd + A^{0,1}, \quad  A^{0,1} \in \Omega^{0,1}(U, \u{r}).
\end{align}

\section{A simple quiver gauge theory} \label{sec:theory}

We are interested in a theory with a single Higgs field $\phi$ that transforms under the gauge groups $\U{r_1}$ and $\U{r_2}$ as follows,
\begin{align}
	\phi \mapsto g_1\phi \, g_2^{-1}, \quad g_1\in \U{r_1}, \quad g_2\in \U{r_2}. \label{eq:Higgs:transformation}
\end{align}
We introduce the local gauge potentials $A_1, A_2$, corresponding to the gauge groups $\U{r_1}$ and $\U{r_2}$ respectively,
\begin{align}
	A_1 \in \Omega^1(U, \u{r_1}), \quad A_2 \in \Omega^1(U, \u{r_2}), \quad U\subset M. \label{eq:gauge:potentials}
\end{align}
The gauge potentials $A_1, A_2$ give rise to terms of Yang--Mills type in our theory. Before we give the static energy functional for our theory, we cast definitions of its ingredients in the geometric language of Section \ref{sec:bundles}.

The gauge groups $\U{r_1}$ and $\U{r_2}$ are identified with the structure groups of two vector bundles $E_1$, $E_2$ on $M$. The ranks of $E_1$, $E_2$ are $r_1$, $r_2$ respectively. The fact that the gauge groups are unitary implies that $E_1$, $E_2$ carry hermitian structures. The gauge potentials correspond to covariant derivatives on $E_1$, $E_2$, which are locally given by
\begin{align}
	D_1 = \d + A_1, \quad D_2 = \d + A_2. \label{eq:D1D2}
\end{align}
We denote the field strengths of $A_1, A_2$ as $F_1, F_2$ respectively,
\begin{align}
	F_1 = \d A_1 + A_1\w A_1, \quad F_2 = \d A_2 + A_2\w A_2, \label{eq:field:strengths}
\end{align}
and these of course agree with the curvatures of $E_1, E_2$.

Having introduced the bundles $E_1, E_2$, we can think of the Higgs field as a homomorphism of vector bundles,
\begin{align}
	\phi \colon E_2 \to E_1. \label{eq:Higgs:homomorphism}
\end{align}
Equivalently, $\phi$ is a section of the bundle $E = E_2^*\otimes E_1$, where $E_2^*$ denotes the dual bundle of $E_2$. The covariant derivative on $E$ is naturally induced from $E_1$, $E_2$,
\begin{align}
	D\phi = \d\phi + A_1\phi - \phi A_2. \label{eq:covariant:D}
\end{align}
Note that the curvature of $E$ acts on $\phi$ as follows,
\begin{align}
	F\phi = D\circ D\phi = F_1\phi - \phi F_2. \label{eq:phi:curvature}
\end{align}

\begin{figure}
\begin{center}
	\includegraphics[scale=0.4]{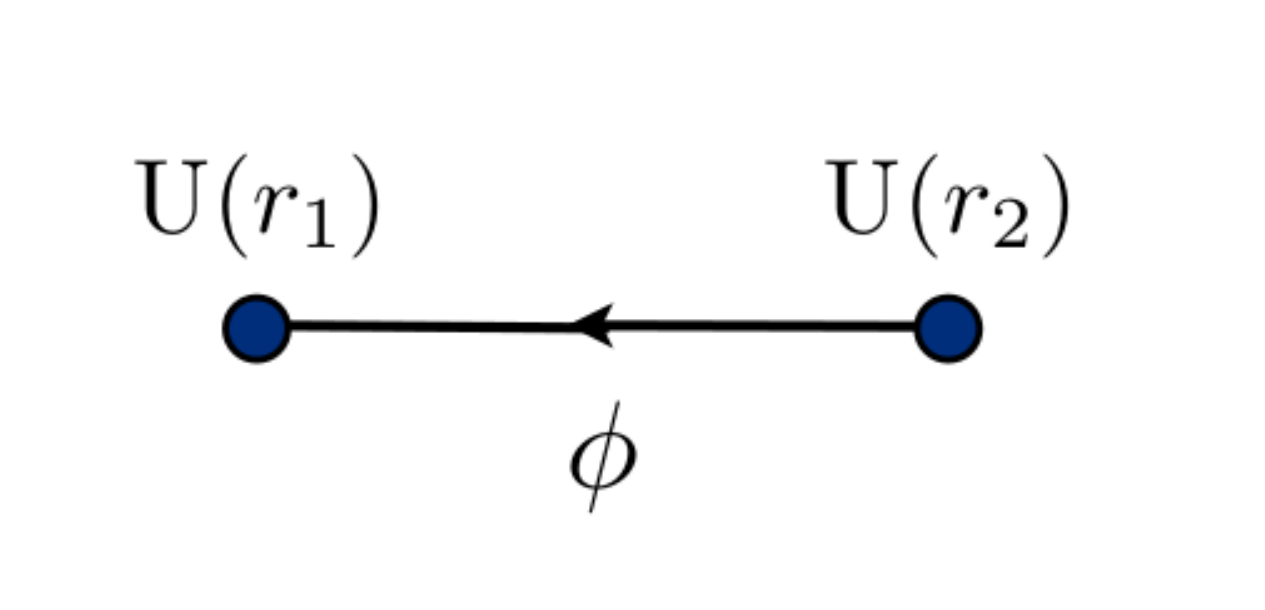}
	\caption{Simple chain quiver diagram for our Yang--Mills--Higgs theory.}
	\label{Fig:quiver}
\end{center}
\end{figure}

The theory we are interested in has the following static energy functional, written in the notation of Section \ref{sec:notation},
\begin{align}
	\curlE(A_1, A_2, \phi) = \mmodsq{F_1} + \mmodsq{F_2} + \mmodsq{D\phi} + V(\phi), 
	\label{eq:YMH:model} 
\end{align}
where
\begin{align}
	V(\phi) = \quart \mmodsq{\phi\phi^\dgr - \tau_1 I_{r_1}} + \quart \mmodsq{\phi^\dgr\phi - \tau_2 I_{r_2}}
\end{align}
is the Higgs field potential from the Introduction. Recall that $\tau_1, \tau_2\in\RR$ are parameters, and $I_{r_1}, I_{r_2}$ denote the identity matrices of rank $r_1, r_2$ respectively. The theory defined by \eqref{eq:YMH:model} is a quiver gauge theory on $M$, whose simple quiver diagram is depicted in Figure \ref{Fig:quiver}.

If one thinks of quivers as a way of classifying gauge theories, then the quiver only fixes the kinetic terms in the corresponding theory. In our case these are the first three terms on the right-hand side of (\ref{eq:YMH:model}), namely $\mmodsq{F_1}$, $\mmodsq{F_2}$, $\mmodsq{D\phi}$. The Higgs field potential $V(\phi)$ is not determined by the quiver, and we have chosen a natural quartic potential, which is commonly encountered in theories that accommodate vortices \cite{Bogomolny, Nielsen:Olesen, Manton:Sutcliffe, Bradlow:line:bundles, GP:direct}. Note that for suitable values of $\tau_1, \tau_2$ one can obtain $\curlE(A_1, A_2, \phi)$ from pure Yang--Mills theory on $M\times\CP{1}$ by equivariant dimensional reduction (cf.~\cite{Popov:quiver, Manton:Rink:nonab, Dolan:Szabo} and references therein). For arbitrary values of $\tau_1, \tau_2$ the Higgs field potential $V(\phi)$ can still be obtained in a natural way if one requires the quiver gauge theory to be supersymmetric, as explained in the Introduction. Note that in the presence of supersymmetry, the functional $\curlE(A_1, A_2, \phi)$ describes only the bosonic part of the theory.

We remark that more complex quivers diagrams than the one above, and the corresponding gauge theories, appear in \cite{Popov:quiver, Popov:double}. In the context of $\SU{2}$-equivariant reductions we can easily obtain chain quivers  with more than two nodes (Fig.~\ref{Fig:quiver}), while with higher-rank reductions we can get much more involved quivers \cite{Dolan:Szabo,Lechtenfeld:SU(3)}. It would be interesting future work to extend the analysis from the present report to those quiver gauge theories.

To conclude this Section, we give the static field equations derived from the energy functional in \eqref{eq:YMH:model},
\begin{align}
	&D * F_1 = \half \left( \phi(*D\phi^\dgr) - (*D\phi)\phi^\dgr \right), \label{eq:YMH:field:1} \\
	&D * F_2 = \half \left( -(*D\phi^\dgr)\phi + \phi^\dgr(*D\phi) \right), \label{eq:YMH:field:2} \\
	&D * D\phi = (\phi\phi^\dgr - \tau) \phi, \label{eq:YMH:field:3}
\end{align}
where $\tau = \frac{\tau_1 + \tau_2}{2}$. Note that 
\begin{align}
	D * F_1 = \d *F_1 + A_1\w*F_1 - *F_1\w A_1,
\end{align}	
and analogously for $F_2$.

\subsection{The BPS equations} \label{sec:BPS:equations}

By a Bogomolny-type argument \cite{Bogomolny} the energy functional \eqref{eq:YMH:model} can be expressed as a sum of positive terms and topological terms\footnote{See the paragraph after \eqref{eq:Chern:char} for what we mean by \textit{topological terms}.}. If the topology is fixed, then $\curlE(A_1, A_2, \phi)$ is minimized by solutions of a set of first oder differential equations, the BPS equations.

Our Bogomolny-type argument is a generalization of the one in \cite{Bradlow:line:bundles}. We start by introducing the following functional,
\begin{align}
	\curlE'(A_1, A_2, \phi)
		&=         4 \mmodsq{F_1^{0,2}} + 4 \mmodsq{F_2^{0,2}} + 2 \mmodsq{D^{0,1}\phi} \nonumber \\
		&\ph{=} 	+ \Mmodsq{\im \Lambda F_1 + \half \phi\phi^\dgr - \frac{\tau_1}{2} I_{r_1} }
					+ \Mmodsq{\im \Lambda F_2 - \half \phi^\dgr\phi + \frac{\tau_2}{2} I_{r_2} } \nonumber \\
	    	&\ph{=} + 2\pi \tau_1\, \C_1(E_1,\omega) - 2\pi \tau_2\, \C_1(E_2,\omega) \nonumber \\
	    	&\ph{=} - 8\pi^2 \Ch_2(E_1, \omega) -8\pi^2 \Ch_2(E_2, \omega), \label{eq:YMH:prime}
\end{align}
and we claim that $\curlE'(A_1, A_2, \phi)$ is equal to the energy functional in \eqref{eq:YMH:model}. To see this, we first inspect one of the terms on the second line of \eqref{eq:YMH:prime},
\begin{align}
	\Mmodsq{\im \Lambda F_1 + \half \phi\phi^{\dgr} - \frac{\tau_1}{2}I_{r_1}} 
	&= \Mmodsq{\Lambda F_1} + \quart \Mmodsq{\phi\phi^{\dgr} - \tau_1 I_{r_1}} \nonumber \\
	&\ph{=} + \int_M \tr{\im\Lambda F_1\phi\phi^{\dgr}} \dvol_M - 2\pi\tau_1\, C_1(E_1,\omega),
\end{align}
where we used
\begin{align}
 \im\int_M \tr{\Lambda F_1} \dvol_M = \im\int_M \tr{F_1}\w \omega^{[d-1]} = 2\pi \C_1(E_1,\omega)\,. \label{eq:trace:F}
\end{align}
We quote the following identity from \cite{Bradlow:line:bundles},
\begin{align}
	\modsq{F_1}\omega^{[d]} 	= \tr{F_1\w F_1}\w \omega^{[d-2]} + \modsq{\Lambda F_1} \omega^{[d]} 
						  + 2(\modsq{F_1^{2,0}} + \modsq{F_1^{0, 2}}) \omega^{[d]}, \label{eq:curvature:squared}
\end{align}
and we refer to \cite{Kobayashi:vector:bundles} for a derivation of this. Noting that $F_1^\dgr = -F_1$ implies $\mmodsq{F_1^{0,2}} = \mmodsq{F_1^{2,0}}$, we obtain
\begin{align}
	4 \mmodsq{F_1^{0,2}} + \Mmodsq{\im \Lambda F_1 + \half \phi\phi^{\dgr} - \frac{\tau_1}{2}I_{r_1}} 
	&= \Mmodsq{F_1} + \quart \Mmodsq{\phi\phi^{\dgr} - \tau_1 I_{r_1}} \nonumber \\
	&\ph{=} + 8\pi^2 \Ch_2(E_1, \omega) - 2\pi\tau_1\, C_1(E_1,\omega) \nonumber \\
	&\ph{=} + \int_M \tr{\im\Lambda F_1\phi\phi^{\dgr}} \dvol_M\,.\label{eq:potential:1}
\end{align}
An analogous analysis can be carried out for the terms in \eqref{eq:YMH:prime} that involve $F_2$. Using \eqref{eq:potential:1} and the corresponding result from that analysis, we arrive at 
\begin{align}
	\curlE'(A_1, A_2, \phi)
	&=	\mmodsq{F_1} + \mmodsq{F_2} + 2 \mmodsq{D^{0,1}\phi} \nonumber \\
	&\ph{=}	+ \quart \Mmodsq{\phi\phi^{\dgr} - \tau_1 I_{r_1}}
			+ \quart \Mmodsq{\phi^{\dgr}\phi - \tau_2 I_{r_2}} \nonumber \\
	&\ph{=}	+ \im \int_M \tr{(\Lambda F_1\phi - \phi \Lambda F_2) \phi^{\dgr}} \dvol_M. \label{eq:YMH:prime:modified}
\end{align}
It remains to identify within $\curlE'(A_1, A_2, \phi)'$ the correct kinetic term for the Higgs field. To this end,
\begin{align}
	&\im \int_M \tr{(\Lambda F_1\phi - \phi \Lambda F_2) \phi^{\dgr}} \dvol_M  \nonumber \\
	&= \im \lpp \Lambda F_1\phi - \phi \Lambda F_2, \phi \rpp  \nonumber \\
	&= \im \lpp \Lambda F^{1,1}\phi, \phi \rpp  \nonumber \\
	&= \im  \lpp \Lambda D^{0,1}D^{1,0}\phi + \Lambda D^{1,0}D^{0,1}\phi, \phi \rpp 
\end{align}
where we used \eqref{eq:phi:curvature} and $\Lambda F = \Lambda F^{1,1}$ in going to the third line. To make further progress we need the generalized K\"ahler identities \cite{Bradlow:line:bundles}, 
\begin{align}
	[\Lambda, D^{1,0}] &=  \im (D^{0,1})^*, \label{eq:gen:Kahler:1} \\
	[\Lambda, D^{0,1}] &= -\im (D^{1,0})^*, \label{eq:gen:Kahler:2}
\end{align}
also known as Nakano identities \cite{Huybrechts}. The operators $(D^{1,0})^*, (D^{0,1})^*$ denote the adjoints of $D^{1,0}, D^{1,0}$, with respect to the scalar product $\lpp\, , \,\rpp$. It follows that
\begin{align}
&\im \int_M \tr{(\Lambda F_1\phi - \phi \Lambda F_2) \phi^{\dgr}} \dvol_M  \nonumber \\
	&\nonumber=  \im \lpp -\im (D^{1,0})^* D^{1,0}\phi + \im (D^{0,1})^*D^{0,1}\phi, \phi \rpp \\
	&=  \mmodsq{D^{1,0}\phi} - \mmodsq{D^{0,1}\phi},
\end{align}
and therefore,
\begin{align}
	\im \int_M \tr{(\Lambda F_1\phi - \Lambda \phi F_2) \phi^{\dgr}} \dvol_M  + 2 \mmodsq{D^{0,1}\phi} = \mmodsq{D\phi}.
\end{align}
Using this in \eqref{eq:YMH:prime:modified}, we finally arrive at $\curlE'(A_1, A_2, \phi)' = \curlE(A_1, A_2, \phi)$, as claimed.

From \eqref{eq:YMH:prime} it is clear that if the topologies of $E_1$, $E_2$ are fixed, then the energy functional $\curlE(A_1, A_2, \phi)$ is minimised by solutions of the following BPS equations,
\begin{align}
	&F_1^{0,2} = 0, \label{eq:BPS:hol:1} \\
	&F_2^{0,2} = 0, \label{eq:BPS:hol:2} \\
	&D^{0,1}\phi = 0, \label{eq:BPS:hol:Higgs} \\
	&\im \Lambda F_1 = \half (\tau_1 I_{r_1} - \phi\phi^\dgr), \label{eq:BPS:1} \\
	&\im \Lambda F_2 = \half (-\tau_2 I_{r_2} + \phi^\dgr\phi) \label{eq:BPS:2}.
\end{align}
For field configurations satisfying these BPS equations, the energy functional receives contributions only from the topological terms, i.e.
\begin{align}
	\curlE(A_1, A_2, \phi) 
	&=_{\text{BPS}}  	 2\pi \tau_1\, \C_1(E_1,\omega) - 2\pi \tau_2\, \C_1(E_2,\omega) \nonumber \\
	&\ph{=_{\text{BPS}}} - 8\pi^2 \Ch_2(E_1, \omega) -8\pi^2 \Ch_2(E_2, \omega)\,.\label{eq:YMBPS}
\end{align}

The mathematical interpretation of \eqref{eq:BPS:hol:1}, \eqref{eq:BPS:hol:2} is that $E_1$, $E_2$ must be holomorphic vector bundles, while \eqref{eq:BPS:hol:Higgs} means that $\phi$ is a holomorphic section of $E_2^*\otimes E_1$, with the holomorphic structure induced from $E_1$, $E_2$. The remaining equations \eqref{eq:BPS:1} and \eqref{eq:BPS:2} are generalizations of the Hermite--Einstein equation \cite{Kobayashi:vector:bundles}, which can be obtained by setting $\phi=0$. Note that \eqref{eq:BPS:1} and \eqref{eq:BPS:2} are also natural extensions of the vortex equations on a Riemann surface \cite{ Manton:Sutcliffe,GP:direct}. By the moduli space of solutions of \eqref{eq:BPS:hol:1}-\eqref{eq:BPS:2} we mean, as usual, the space of solutions modulo gauge transformations. We remark that this moduli space is obviously contained in the moduli space of holomorphic structures on $E_1$, $E_2$.

By equivariant dimensional reduction the BPS equations \eqref{eq:BPS:hol:1}-\eqref{eq:BPS:2} arise naturally from the Donaldson--Uhlenbeck--Yau (DUY) equations \cite{donaldson1,donaldson2,uhlenbeckyau1,uhlenbeckyau2} for pure Yang--Mills theory on 
$M\times \CP{1}$. The first order DUY equations are
\begin{align}
 \curlF^{2,0} = \curlF^{0,2} = 0, \quad \text{and} \quad \curlF^{1,1} \w *\Omega = 0,
\end{align}
where $\Omega$ is the K\"ahler form on $M\times \CP{1}$ and $\mathcal{F}$ denotes the Yang--Mills field strength. The DUY equations imply the second order Yang--Mills equation $D * \curlF=0$, but the converse is not true. Note that if $M\times \CP{1}$ has complex dimension two, the DUY equations are equivalent to the standard self-duality condition $\curlF = *\curlF$. Deriving the BPS equations \eqref{eq:BPS:hol:1}-\eqref{eq:BPS:2} from the DUY equations on $M\times \CP{1}$ has the characteristic that the values of $\tau_1$ and $\tau_2$ are fixed uniquely by the precise details of the reduction \cite{Dolan:2010ur}.

\subsection{Lower energy bounds}

In order to understand how the vacuum of the theory defined by the energy functional $\curlE(A_1, A_2, \phi)$ is related to the lowest BPS state, we derive two lower bounds for $\curlE(A_1, A_2, \phi)$. The first bound will hold in general and we will refer to it as the \textit{a priori} bound (cf.~\cite{Manton:Rink:nonab}). The second bound will apply to solutions of the BPS equations \eqref{eq:BPS:hol:1}-\eqref{eq:BPS:2} and, of course, will be greater or equal to the a priori bound.

For the a priori bound, we start with the following estimate,
\begin{align}
	\curlE(A_1, A_2, \phi) \ge \quart \mmodsq{\phi\phi^\dgr - \tau_1 I_{r_1}} + \quart \mmodsq{\phi^\dgr\phi - \tau_2 I_{r_2}}. \label{eq:priori:estimate}
\end{align}
The (implicit) traces on the right-hand side can be expanded and rearranged,
\begin{align}
	&\quart \tr{\phi\phi^\dgr - \tau_1 I_{r_1}}^2 + \quart \tr{\phi^\dgr\phi - \tau_2 I_{r_2}}^2 \nonumber \\
	&\nonumber = \half \left(\frac{\tau_1^2 r_1 + \tau_2^2 r_2}{2} - 2\tau\, \tr{\phi \phi^\dgr} + \tr{\phi\phi^\dgr\phi\phi^\dgr} \right) \\
	&= \frac{\tau_1^2r_1 + \tau_2^2r_2 - 2 \tau^2r_1}{4} + \half \tr{\tau I_{r_1} - \phi\phi^\dgr}^2, \label{eq:traces}
\end{align}
where $\tau = \frac{\tau_1 + \tau_2}{2}$, as before. We therefore obtain
\begin{align}
	\curlE(A_1, A_2, \phi) \ge \frac{\vol{M}}{4} (\tau_1^2r_1 + \tau_2^2r_2 - 2 \tau^2r_1), \label{eq:YMH:lower}
\end{align}
and analogously
\begin{align}
	\curlE(A_1, A_2, \phi) \ge \frac{\vol{M}}{4} (\tau_1^2r_1 + \tau_2^2r_2 - 2 \tau^2r_2). \label{eq:YMH:lower:r2}
\end{align}

From now on we assume $r_2\ge r_1$, which presents no loss of generality. Note that for $r_2\ge r_1$ the lower bound \eqref{eq:YMH:lower} is stricter than \eqref{eq:YMH:lower:r2}, and, henceforth, whenever we speak of \textit{the} a priori bound, we shall mean \eqref{eq:YMH:lower}.
Note also that for $r_2\ge r_1$ the right-hand side of \eqref{eq:YMH:lower} is non-negative,
\begin{align}
	(\tau_1^2r_1 + \tau_2^2r_2 - 2 \tau^2r_1) \ge (\tau_1^2 + \tau_2^2 - 2 \tau^2)r_1 = \half (\tau_1 - \tau_2)^2 r_1.
\end{align}
Particularly, the a priori bound is strictly positive unless $r_1 = r_2$ and $\tau_1 = \tau_2$. This implies that, for general values of  $r_1$, $r_2$, $\tau_1$, $\tau_2$, the theory defined by $\curlE(A_1, A_2, \phi)$ has non-vanishing vacuum energy.

To derive the bound that applies to BPS states, we need the following estimates,
\begin{align}
	\mmodsq{F_1} \ge \frac{1}{d}\mmodsq{\Lambda F_1^{1,1}}, \quad \mmodsq{F_2} \ge \frac{1}{d} \mmodsq{\Lambda F_2^{1,1}},
\end{align}
which we establish in appendix \ref{app:curvature}. We can thus estimate $\curlE(A_1, A_2, \phi)$ as follows,
\begin{align}
	\curlE(A_1, A_2, \phi)
	 	&\ge \mmodsq{F_1}  + \mmodsq{F_2} + \quart \mmodsq{\phi\phi^\dgr - \tau_1 I_{r_1}} + \quart \mmodsq{\phi^\dgr\phi - \tau_2 I_{r_2}} \\
		&\ge \frac{1}{d}\mmodsq{\Lambda F_1^{1,1}}  + \frac{1}{d} \mmodsq{\Lambda F_2^{1,1}} \nonumber \\
		&\ph{\ge} + \quart \mmodsq{\phi\phi^\dgr - \tau_1 I_{r_1}} + \quart \mmodsq{\phi^\dgr\phi - \tau_2 I_{r_2}}.
\end{align}
Next we rewrite $\Lambda F_1^{1,1}$, $\Lambda F_2^{1,1}$ using the BPS equations \eqref{eq:BPS:1}, \eqref{eq:BPS:2},
\begin{align}
	\curlE(A_1, A_2, \phi) & \ge_{\text{BPS}}  \frac{d+1}{d} \left(\quart \mmodsq{\phi\phi^\dgr - \tau_1 I_{r_1}} + \quart \mmodsq{\phi^\dgr\phi - \tau_2 I_{r_2}} \right).
\end{align}
Rearranging traces as in \eqref{eq:traces}, we obtain the lower bound
\begin{align}
	\curlE(A_1, A_2, \phi) \ge_{\text{BPS}} \frac{\vol{M}}{4} \frac{d+1}{d} (\tau_1^2r_1 + \tau_2^2r_2 - 2 \tau^2r_1). \label{eq:BPS:lower}
\end{align}
For general values of  $r_1$, $r_2$, $\tau_1$, $\tau_2$ there is a strictly positive gap between this lower bound for BPS states and the a priori bound in \eqref{eq:YMH:lower}. This has the dramatic consequence that if the a priori bound is attained by the vacuum of our theory \eqref{eq:YMH:model}, then the vacuum is not a BPS state. 

We can in fact be more explicit about the vacua of \eqref{eq:YMH:model}. The lower bound \eqref{eq:YMH:lower} is attained by field configurations that satisfy
\begin{align}
	&F_1 = 0, \label{eq:curvature:1} \\
	&F_2 = 0, \label{eq:curvature:2} \\
	&D\phi = 0, \label{eq:constant:Higgs} \\
	&\phi = \sqrt{\tau} \left(\begin{array}{cc} I_{r_1} & 0 \end{array}\right). \label{eq:global:Higgs}
\end{align}
If we let $E_1$, $E_2$ be the trivial bundles of ranks $r_1$, $r_2$ respectively, then it is topologically consistent to choose $A_1 = 0$, $A_2 = 0$, which solve \eqref{eq:curvature:1}, \eqref{eq:curvature:2}. Also because of the triviality of $E_1$, $E_2$ the choice of Higgs field in \eqref{eq:global:Higgs} is globally meaningful. This proves the existence of a vacuum state for which the a priori lower bound is attained
\begin{equation}
	\curlE_{\text{vac}} = \frac{\vol{M}}{4} (\tau_1^2r_1 + \tau_2^2r_2 - 2 \tau^2r_1). \label{eq:VacEnergy}
\end{equation} 
By the reasoning at the end of the previous paragraph this vacuum state is not a BPS state, and the BPS bound (\ref{eq:BPS:lower}) can be rewritten as
\begin{equation}
	\curlE(A_1, A_2, \phi) \ge_{\textit{BPS}} \frac{d+1}{d} \curlE_{\text{vac}} .\label{BPSbound}
\end{equation} 

The vacuum state breaks supersymmetry due to the presence of non-vanishing $D$-terms, and it also breaks the gauge symmetry $\U{r_1}\times \U{r_2}$ to a diagonal $\U{r_1}$ times $\U{r_2 - r_1}$. To be more specific, the vacuum value of the Higgs field,
\begin{align}
	\phi = \sqrt{\tau}  \left(\begin{array}{cc} I_{r_1} & 0 \end{array}\right),
\end{align}
is invariant under gauge transformations
\begin{align}
	\phi \mapsto g_1 \phi g_2^{-1},
\end{align}
with $g_1\in \U{r_1}$ and
\begin{align}
	g_2 = \left(\begin{array}{cc} g_1^{-1} & \\  & g' \end{array}\right),
\end{align}
where $g' \in \U{r_2 - r_1}$.

\section{Topologically non-trivial non-BPS solutions} \label{sec:non-BPS}

Having established that there is generally a non-trivial gap between the a priori bound \eqref{eq:YMH:lower} and the BPS bound \eqref{eq:BPS:lower} for our energy functional $\curlE(A_1, A_2, \phi)$, we now study properties of such solutions of the field equations \eqref{eq:YMH:field:1}-\eqref{eq:YMH:field:3} whose energies lie in this gap. To this end, we consider the following set of equations,
\begin{align}
	&F_1^{0,2} = 0, \label{eq:hol:bundle:1} \\
	&F_2^{0,2} = 0, \label{eq:hol:bundle:2} \\
	&\phi = \sqrt{\tau} \left(\begin{array}{cc} I_{r_1} & 0 \end{array}\right), \label{eq:trivial:Higgs} \\
	&\im\Lambda F_1 = \half (\tau_1 - \tau) I_{r_1}, \label{eq:HE:1} \\
	&\im\Lambda F_2 = \half \left(\begin{array}{ccc} (\tau-\tau_2) I_{r_1} & 0 & 0 \\ 0 & -\tau_2 I_{(r_2-r_1-k)} & 0 \\ 0 & 0 & 0 \end{array}\right), \label{eq:HE:2}
\end{align}
where $k \in \{0, \dots, r_2-r_1\}$, and the bottom-right entry on the right-hand side of \eqref{eq:HE:2} is the $k\!\times\!k$ zero matrix. Note that the above equations make sense globally only if $E_2$ decomposes as follows,
\begin{align}
	E_2 = E_1 \oplus E'  \oplus E_0, \label{eq:E2:decomp}
\end{align}
where $\rk(E') = (r_2-r_1-k)$ and $\rk(E_0)= k$. Moreover, equations \eqref{eq:HE:1} and \eqref{eq:HE:2} can be solved precisely if $E_1$, $E'$, and $E_0$ are Hermite--Einstein. As in Section \ref{sec:BPS:equations} we refer to \cite{Kobayashi:vector:bundles} for a definition of Hermite--Einstein vector bundles and for the solution theory of \eqref{eq:HE:1}, \eqref{eq:HE:2}. Later in this Section we will look at solutions of \eqref{eq:HE:1}, \eqref{eq:HE:2} in the case where $E_1$ and $E_2$ decompose into line bundles.

In the next Subsection we will verify that solutions of \eqref{eq:hol:bundle:1}-\eqref{eq:HE:2} also satisfy the field equations \eqref{eq:YMH:field:1}-\eqref{eq:YMH:field:3}. However, solutions of \eqref{eq:hol:bundle:1}-\eqref{eq:HE:2} do not solve the BPS equations unless $k=0$. This is immediately clear upon comparing \eqref{eq:HE:2} with \eqref{eq:BPS:2}. 

Using \eqref{eq:trace:F} one obtains the following first Chern numbers for solutions of \eqref{eq:hol:bundle:1}-\eqref{eq:HE:2},
\begin{align}
	\C_1(E_1,\omega) &= \frac{\vol{M}}{4\pi} (\tau_1 -\tau) r_1, \\
	\C_1(E_2,\omega) &= \frac{\vol{M}}{4\pi} \left( (\tau-\tau_2) r_1 -\tau_2(r_2-r_1-k)\right).
\end{align}
This justifies our terminology to refer to solutions of \eqref{eq:hol:bundle:1}-\eqref{eq:HE:2} as \textit{topologically non-trivial non-BPS states}. Different values of $k \in\{1, \dots, r_2-r_1\}$ correspond to different such non-BPS states. The above Chern numbers vanish for $\tau_1=\tau_2$ and $k = r_2-r_1$, which corresponds to the topologically trivial, non-BPS vacuum of  $\curlE(A_1, A_2, \phi)$.

\subsection{Solving the field equations}

We now check that solutions of \eqref{eq:hol:bundle:1}-\eqref{eq:HE:2} also solve the field equations \eqref{eq:YMH:field:1}-\eqref{eq:YMH:field:3}. Note that the decomposition \eqref{eq:E2:decomp} implies
\begin{align}
	A_2 = \left(\begin{array}{ccc} A_1 & 0 & 0 \\ 0 & A' & 0 \\ 0 & 0 & A_0 \end{array}\right),
\end{align}
where $A'$, $A_0$ are the connections on $E'$, $E_0$ respectively. Combining the decomposition of $A_2$ with \eqref{eq:trivial:Higgs}, it follows that
\begin{align}
	D\phi = 0.
\end{align}
This, again in combination with \eqref{eq:trivial:Higgs}, shows that \eqref{eq:YMH:field:3} is satisfied. Furthermore, the field equations \eqref{eq:YMH:field:1}, \eqref{eq:YMH:field:2} reduce to
\begin{align}
	&D * F_1 = 0, \\
	&D * F_2 = 0,
\end{align}
which the following Lemma serves to verify.

\newtheorem{Lemma}{Lemma}
\begin{Lemma} \label{Lemma}
Let $E$ be a vector bundle over a K\"ahler manifold, and let $F$ be the curvature of $E$. Assume $F$ satisfies the following equations,
\begin{align}
	&F^{2,0} = 0 = F^{0,2}, \\
	&\Lambda F = c I,
\end{align}
where $c\in\CC$ is a constant and $I\colon E \to E$ is the identity map. Then $D*F = 0$.
\end{Lemma}

We defer a proof of Lemma \ref{Lemma} to appendix \ref{app:Lemma}. Note that a version of this Lemma already featured informally in \cite{Park}.

\subsection{The energy ladder} \label{sec:ladder}

In order to evaluate our energy functional \eqref{eq:YMH:model}, we specialize equations \eqref{eq:HE:1} and \eqref{eq:HE:2} as follows,
\begin{align}
	&F_1 = -\ihalf \frac{\omega}{d} (\tau_1 - \tau) I_{r_1}\,, \label{eq:F1:omega} \\
	&F_2 = -\ihalf \frac{\omega}{d} \left(\begin{array}{ccc} (\tau-\tau_2) I_{r_1} & 0 & 0 \\ 0 & -\tau_2 I_{(r_2-r_1-k)} & 0 \\ 0 & 0 & 0 \end{array}\right). \label{eq:F2:omega}
\end{align}
From this it follows straightforwardly that
\begin{align}
	& \Mmodsq{F_1} = \frac{\vol{M}}{4d}(\tau_1 - \tau)^2 r_1, \\
	&\Mmodsq{F_2} = \frac{\vol{M}}{4d}\left((\tau - \tau_2)^2 r_1 + \tau_2^2(r_2-r_1-k)\right),
\end{align}
and hence,
\begin{align}
	\curlE(A_1, A_2, \phi) 	&\nonumber= \frac{\vol{M}}{8d}(\tau_1 - \tau_2)^2 r_1 + \frac{\vol{M}}{4d} \tau_2^2(r_2-r_1-k) \\
		  				&\label{eq:NonBPSenergy}\ph{=} + \frac{\vol{M}}{4} (\tau_1^2r_1 + \tau_2^2r_2 - 2 \tau^2r_1),
\end{align}
where we used \eqref{eq:traces}. From this expression for $\curlE(A_1, A_2, \phi)$ it is clear that the energy gap $\Delta$ between solutions of \eqref{eq:hol:bundle:1}-\eqref{eq:HE:2} with subsequent values of $k$ is
\begin{equation}
\label{eq:Delta}	\Delta = \frac{\vol{M}}{4d} \tau_2^2 .
\end{equation}
Since $\Delta$ is independent of $k$, we call the set of solutions of \eqref{eq:hol:bundle:1}-\eqref{eq:HE:2} a \textit{ladder} of topologically non-trivial non-BPS states, which justifies the title of this report.

\begin{figure}[t!]
  \begin{center} 
	\includegraphics[scale=0.4]{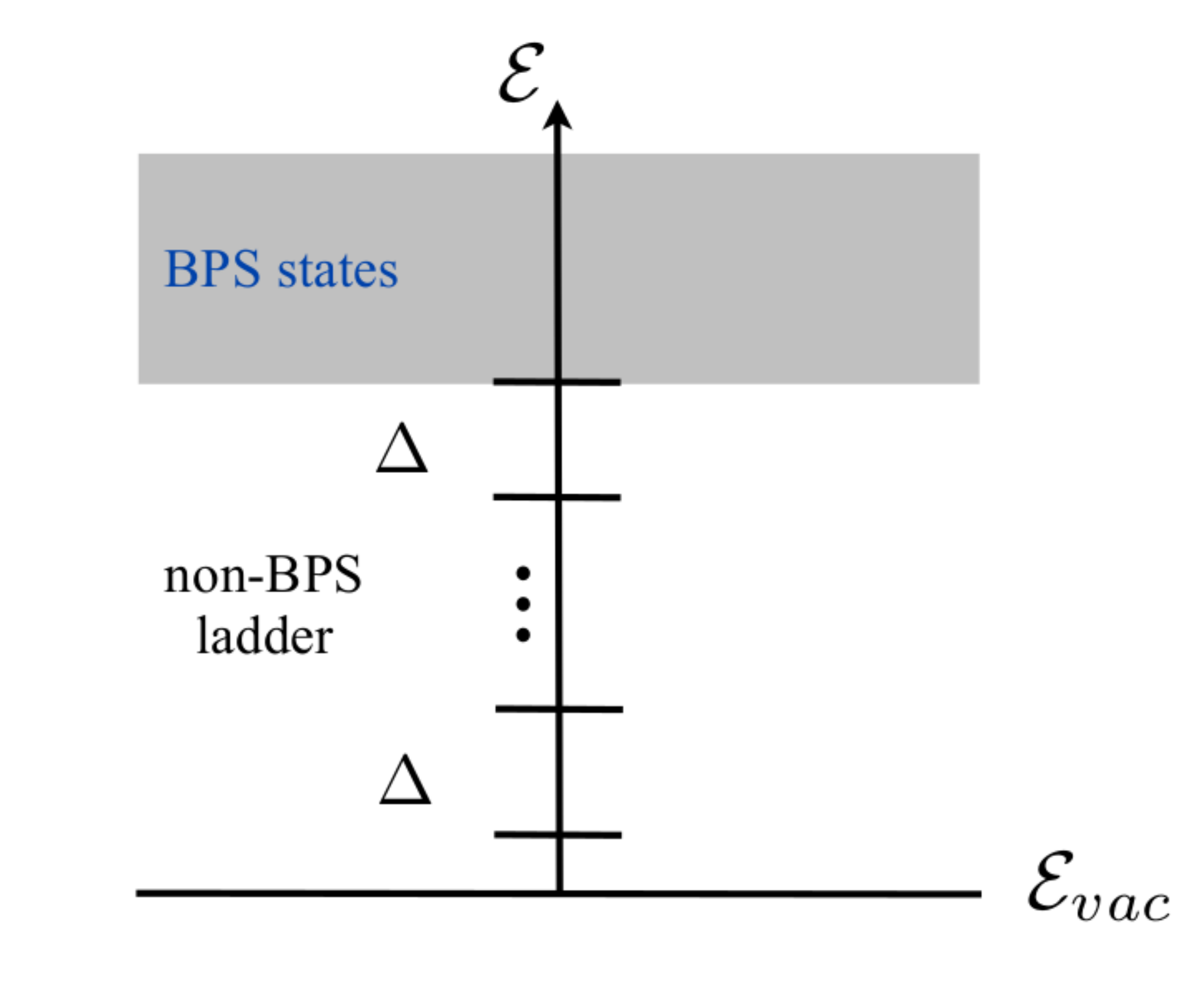}
	\caption{\label{fig:1} {\em Energy ladder of non-BPS, topologically non-trivial states.}}
   \end{center}
\end{figure}

We can rewrite (\ref{eq:NonBPSenergy}) in the following, more illuminating equivalent ways
\begin{align}
	\curlE(A_1, A_2, \phi)	&= \curlE_{\text{vac}}+ \frac{\vol{M}}{8d}(\tau_1 - \tau_2)^2 r_1+\Delta (r_2-r_1-k)\,,\label{eq:ladder:1} \\
						&=\curlE_{\text{BPS}} - \Delta\,k\,, \label{eq:ladder:2}
\end{align}
where we used \eqref{eq:VacEnergy}, and we introduced $\curlE_{\text{BPS}}$ to denote the expression on the right-hand side of the lower energy bound \eqref{eq:BPS:lower}, i.e.
\begin{align}
	\curlE_{\text{BPS}} = 	\frac{\vol{M}}{4d} (d+1) (\tau_1^2r_1 + \tau_2^2r_2 - 2 \tau^2r_1).
\end{align}
Figure \ref{fig:1} illustrates equations \eqref{eq:ladder:1}, \eqref{eq:ladder:2}: The lowest non-BPS state of the ladder, corresponding to $k=r_2-r_1$, has strictly greater energy than the vacuum (provided $\tau_1\ne\tau_2$). The energy levels of the more energetic states, corresponding to $k \in\{0,1,...,r_2-r_1-1\}$, are equidistant, with gap $\Delta$. When $k=0$, the lowest energy BPS state is attained; it is straightforward to see that in this case equations \eqref{eq:trivial:Higgs}-\eqref{eq:HE:2} describe a BPS state.

\subsection{Existence of non-BPS solutions}

In constructing special solutions of the BPS equations \eqref{eq:BPS:hol:1}-\eqref{eq:BPS:2} it is a standard trick to assume that $E_1, E_2$ decompose into line bundles. In this subsection we apply the same trick to equations \eqref{eq:hol:bundle:1}-\eqref{eq:HE:2}.

We first recall a result from \cite{Bradlow:line:bundles}. Let $L$ be a complex line bundle on $M$, equipped with a hermitian structure, and denote as  $F$ the curvature of $L$. Since $L$ is a line bundle, we have $F=\d A$, where $A$ is a unitary connection on $L$. The Hermite--Einstein condition (cf.~\cite{Kobayashi:vector:bundles}),
\begin{align}
	\im\Lambda F = \chi, \label{eq:HE:line:bundle}
\end{align}
where $\chi \in \RR$ is a constant, can be solved for $A$ precisely if 
\begin{align}
	\C_1(L, \omega) = \frac{\vol{M}}{2\pi}\chi
\end{align}
is an integer. Furthermore, the moduli space of solutions, up to \U{1}-gauge transformations, is in 1-1 correspondence with equivalence classes of holomorphic structures on $L$. If $d=1$, i.e.~if $M$ is a Riemann surface, then this moduli space agrees with the Jacobian of $M$.

Provided the combination of $\tau_1$, $\tau_2$, and $\vol{M}$ is chosen such that
\begin{align}
	&\frac{\vol{M}}{4\pi}(\tau_1-\tau) \in \ZZ, \\
	&\frac{\vol{M}}{4\pi} \tau_2 \in \ZZ,
\end{align}
then solutions of \eqref{eq:HE:1}, \eqref{eq:HE:2} can be constructed as follows: Let $L_1, L'$ be holomorphic line bundles on $M$ whose respective curvatures $F_{L_1}, F_{L'}$ satisfy
\begin{align}
	&\im\Lambda F_{L_1} =  \half (\tau_1 - \tau) , \label{eq:HE:line:bundle:1} \\
	&\im\Lambda F_{L'} =  -\half \tau_2 , \label{eq:HE:line:bundle:2}
\end{align}
and set
\begin{align}
	E_1 = \underbrace{L_1 \oplus \dots \oplus L_1}_{r_1\text{ times}}, \quad
	E' = \underbrace{L' \oplus \dots \oplus L'}_{r_2-r_1-k\text{ times}}.
\end{align}
As before we set $E_2 = E_1 \oplus E'  \oplus E_0$, from which it follows that \eqref{eq:trivial:Higgs} can be solved, and \eqref{eq:hol:bundle:1},  \eqref{eq:hol:bundle:2} hold trivially since $E_1$ and $E'$ are holomorphic bundles.

\subsection{The case $d=1$}

The case $d=1$, i.e.~where $M$ is a Riemann surface, is of particular interest: In this case BPS solutions of \eqref{eq:BPS:hol:1}-\eqref{eq:BPS:2} are referred to as vortices, which have received ample attention in the literature \cite{ Popov:nonab, Manton:Rink:nonab,Nielsen:Olesen,  Manton:Sutcliffe,GP:direct, Popov:int}.\footnote{The term vortex is occasionally also used to refer to BPS solutions in higher dimensions \cite{ Popov:quiver,Bradlow:line:bundles}.} If $r_1r_2=1$, we speak of abelian vortices; and vortices are non-abelian for $r_1r_2>1$. This is because the structure group of $E_2^*\otimes E_1$ is necessarily abelian if $r_1r_2=1$, but generally non-abelian for $r_1r_2>1$. 

The construction from the previous subsection obviously applies to the case $d=1$. Nonetheless it is worthwhile noting that equations \eqref{eq:HE:line:bundle:1}-\eqref{eq:HE:line:bundle:2} can be replaced with
\begin{align}
	&F_{L_1} =  -\ihalf (\tau_1 - \tau) \omega, \\
	&F_{L'} =  \ihalf \tau_2 \omega.
\end{align}
More generally, on a Riemann surface $M$, equations \eqref{eq:HE:1}, \eqref{eq:HE:2} are equivalent to \eqref{eq:F1:omega}, \eqref{eq:F2:omega} with $d=1$.

For $d=2$ various authors have studied solutions to the BPS equations \eqref{eq:BPS:hol:1}-\eqref{eq:BPS:2} in the context of equivariant dimensional reductions \cite{Popov:2003xg,Lechtenfeld:2007st}. In the case of abelian gauge groups the reduced equations are usually called Seiberg--Witten monopole equations \cite{Witten:1994cg}. On $M=\mathbb{R}^4$ the only finite action solution to the Seiberg--Witten equations is the trivial solution. By introducing a non-commutative deformation of $\mathbb{R}^4$ non-trivial solutions which are regular and have finite energy can be obtained. These non-trivial solutions can be interpreted as D-branes. In this report we circumvented the triviality arguments without introducing non-commutative deformations by choosing the background manifold $M$ to be compact and of finite area.

\section{Acknowledgements}

This work was initiated during NAR's PhD research. The authors wish to thank David Tong and Nick Manton for useful discussions.
DD is grateful for support through the European Research Council Advanced Grant No. 247252, ``Properties and Applications of the Gauge/Gravity Correspondence''.

\appendix

\section{Curvature estimates} \label{app:curvature}

In this appendix we establish the estimate
\begin{align}
	\mmodsq{F^{1,1}}  \ge \frac{1}{d} \mmodsq{\Lambda F^{1,1}},
\end{align}
where $F\in\Omega^2(M,\u{r})$, $r\in\NN$. Of course, in the main body of this report we apply this estimate in the situation where $F$ is the curvature of a vector bundle over $M$.

Our first step is to decompose the 2-form $F$ according to \eqref{eq:form:decomp},
\begin{align}
	F = F^{2,0} + F^{1,1} + F^{0,2},
\end{align}
which implies for the modulus of $F$,
\begin{align}
	\mmodsq{F} = \mmodsq{F^{2,0}} + \mmodsq{F^{1,1}} + \mmodsq{F^{0,2}}.
\end{align}
The $(1,1)$-component can be further decomposed into a part proportional to the K\"ahler form $\omega$ and an orthogonal part,
\begin{align}
	&F^{1,1} = F_{\omega}^{1,1} + F_{\perp}^{1,1}, \label{eq:F11decomposition}\\
	&(\omega, F_{\perp}^{1,1}) = 0.
\end{align}
Hence, for an arbitrary function $\al \in \Omega^0(M,\u{r})$,
\begin{align}
	(\al, \Lambda F_{\perp}^{1,1}) 	&= (L\al, F_{\perp}^{1,1}) \\
							&= \al(\omega, F_{\perp}^{1,1}) \\
							&= 0,
\end{align}
i.e.~$\Lambda F_{\perp}^{1,1} = 0$. Note also that
\begin{align}
	(\omega,\omega) \dvol_M = \omega \w *\omega = d\, \dvol_M,
\end{align}
and therefore $\Lambda\omega = d$. Combining this with $\Lambda F_{\perp}^{1,1} = 0$, it follows that
\begin{align}
	F_{\omega}^{1,1} = \frac{1}{d} (\Lambda F^{1,1}) \omega.
\end{align}
Altogether we can now write for the $(1,1)$-component of $F$,
\begin{align}
	\mmodsq{F^{1,1}} 	&= \mmodsq{F_{\omega}^{1,1}} + \mmodsq{F_{\perp}^{1,1}} \\
					&= \frac{1}{d^2} \int_M \modsq{\Lambda F^{1,1}} (\omega,\omega) \dvol_M + \mmodsq{F_{\perp}^{1,1}} \\
					&= \frac{1}{d} \mmodsq{\Lambda F^{1,1}} + \mmodsq{F_{\perp}^{1,1}}, \label{eq:curvature:11:modulus}
\end{align}
yielding the desired estimate.

\section{Proof of Lemma \ref{Lemma}} \label{app:Lemma}

The first assumption in the Lemma implies $F = F^{1,1}$. Therefore the Bianchi identity reads
\begin{align}
	0 = DF^{1,1} = D^{1,0} F^{1,1} + D^{0,1} F^{1,1}.
\end{align}
Note that the two terms on the right-hand side have different bi-degrees and therefore must vanish separately, i.e.
\begin{align}
	&D^{1,0} F^{1,1} = 0, \label{eq:Bianchi:10} \\
	&D^{0,1} F^{1,1} = 0. \label{eq:Bianchi:01}
\end{align}
Next, consider the action of the adjoint covariant derivative $D^*$ on $F$,
\begin{align}
	D^*F &= D^*F^{1,1} \\
		 &= (D^{1,0})^*F^{1,1} + (D^{0,1})^*F^{1,1} \\
		 &= \im [\Lambda, D^{0,1}] F^{1,1} - \im [\Lambda, D^{1,0}] F^{1,1} \\
		 &= -\im D^{0,1}\Lambda F^{1,1} + \im D^{1,0} \Lambda F^{1,1},
\end{align} 
where we used the generalized K\"ahler identities \eqref{eq:gen:Kahler:1}, \eqref{eq:gen:Kahler:2} in going to the third line, and \eqref{eq:Bianchi:10}, \eqref{eq:Bianchi:01} in going to the last. Now, by the second assumption in the Lemma, $\Lambda F^{1,1} = c I$, and therefore, 
\begin{align}
	D^*F = -\im c \underbrace{D^{0,1} I}_{0} + \im c \underbrace{D^{1,0} I}_{0} = 0.
\end{align}
Since $D^*F = - * D * F$, we obtain the desired result $D * F = 0$.

\bibliographystyle{utphys-new}  
\clearpage 
\addcontentsline{toc}{section}{References}
\bibliography{main}

\end{document}